\documentclass[pra, aps,onecolumn,10pt,showpacs,groupedaddress,superscriptaddress,floatfix,notitlepage]{revtex4-2}
\usepackage{amsmath,amsfonts,amssymb,graphics,graphicx,epsfig,color,times}%,bbm}
\usepackage[utf8x]{inputenc}
\usepackage{color}
\usepackage{bbm, dsfont} 
\usepackage{subfigure}
\usepackage{hyperref}
\usepackage{mathrsfs}
\usepackage{verbatim}

\begin{document}
\title{Superior dark-state cooling via nonreciprocal couplings in trapped atoms}

\author{Chun-Che Wang}
\thanks{These two authors contributed equally}
\email{david23203649@gmail.com}
\affiliation{Institute of Atomic and Molecular Sciences, Academia Sinica, Taipei 10617, Taiwan}

\author{Yi-Cheng Wang}
\thanks{These two authors contributed equally}
\email{r09222006@ntu.edu.tw}
\affiliation{Department of Physics, National Taiwan University, Taipei 10617, Taiwan}
\affiliation{Institute of Atomic and Molecular Sciences, Academia Sinica, Taipei 10617, Taiwan}

\author{Chung-Hsien Wang}
\affiliation{Department of Physics, National Taiwan University, Taipei 10617, Taiwan}
\affiliation{Institute of Atomic and Molecular Sciences, Academia Sinica, Taipei 10617, Taiwan}

\author{Chi-Chih Chen}
\affiliation{Institute of Atomic and Molecular Sciences, Academia Sinica, Taipei 10617, Taiwan}

\author{H. H. Jen}
\email{sappyjen@gmail.com}
\affiliation{Institute of Atomic and Molecular Sciences, Academia Sinica, Taipei 10617, Taiwan}
\affiliation{Physics Division, National Center for Theoretical Sciences, Taipei 10617, Taiwan}

\date{\today}
\renewcommand{\r}{\mathbf{r}}
\newcommand{\f}{\mathbf{f}}
\renewcommand{\k}{\mathbf{k}}
\def\p{\mathbf{p}}
\def\q{\mathbf{q}}
\def\bea{\begin{eqnarray}}
\def\eea{\end{eqnarray}}
\def\ba{\begin{array}}
\def\ea{\end{array}}
\def\bdm{\begin{displaymath}}
\def\edm{\end{displaymath}}
\def\red{\color{red}}
\pacs{}
\begin{abstract}
Cooling the trapped atoms toward their motional ground states is key to applications of quantum simulation and quantum computation. By utilizing nonreciprocal couplings between constituent atoms, we present an intriguing dark-state cooling scheme in $\Lambda$-type three-level structure, which is shown superior than the conventional electromagnetically-induced-transparency cooling in a single atom. The effective nonreciprocal couplings can be facilitated either by an atom-waveguide interface or a free-space photonic quantum link. By tailoring system parameters allowed in dark-state cooling, we identify the parameter regions of better cooling performance with an enhanced cooling rate. We further demonstrate a mapping to the dark-state sideband cooling under asymmetric laser driving fields, which shows a distinct heat transfer and promises an outperforming dark-state sideband cooling assisted by collective spin-exchange interactions. 
\end{abstract}
\maketitle
%%%%%%%%%%%%%%%%%%%%%%%%%%%%%%%%%%%%%%%%%%%%%%%%%%%%%%%%%%%%%%%%%% Introduction 150 words
\section{Introduction} 

Cooling the trapped atoms toward their motional ground states engages a series of improving laser cooling techniques from Doppler to Sisyphus and subrecoil cooling schemes \cite{Cohen2011}. Upon approaching the motional ground state of atoms, the linewidth or the spontaneous decay rate of the transition puts a limit on the temperature that these cooling schemes can achieve. Their ultimate performance depends on the balance between cooling and heating mechanisms, where the fluctuations of spontaneously emitted and rescattered photons give rise to the constraint that forbids further cooling. In one of the subrecoil cooling platforms, the resolved sideband cooling \cite{Diedrich1989, Cirac1992, Monroe1995, Roos1999, Zhang2021} in the Lamb-Dicke regime suppresses the carrier transition and specifically excites the transition with one phononic quanta less. Effectively, it moves the atoms toward the zero-phonon state via spontaneous emissions, and essentially, the steady-state phonon occupation is determined by a squared ratio of the spontaneous emission rate over the trapping frequency, which can be much smaller than one.    

Alternatively, the dark-state sideband cooling \cite{Morigi2000, Roos2000, Morigi2003, Retzker2007, Cerrillo2010, Albrecht2011, Zhang2012, Zhang2014, Kampschulte2014, Lechner2016, Scharnhorst2018, Jordan2019, Feng2020, Qiao2021, Huang2021, Zhang2021_2} utilizes $\Lambda$-type three-level structure of atoms with two laser fields operating on two different hyperfine ground states and one common excited state. This leads to an effective dark-state picture owing to quantum interference between two ground states and forms an asymmetric absorption profile under the two-photon resonance condition with a large detuning. This profile allows a narrow transition in the resolved sideband to cool down the atoms by removing one phonon, similar to the sideband cooling scheme with a two-level structure. In contrast to the two-level structure, the effective laser drivings and decay rates in dark-state sideband cooling can be tunable, which makes it a flexible and widely-used cooling scheme in ultracold atoms. 

Recently, a novel scheme of using nonreciprocal couplings in optomechanical systems \cite{Xu2019, Lai2020} manifests motional refrigeration, which gives insights in heat transfer by utilizing unequal decay channels. This coupling can be facilitated as well in a nanophotonics platform of atom-waveguide interface \cite{Chang2018, Corzo2019, Sheremet2022} which allows quantum state engineering via mediating collective nonreciprocal couplings between atoms \cite{Gardiner1993, Carmichael1993, Stannigel2012, Luxmoore2013, Ramos2014, Arcari2014, Mitsch2014, Pichler2015, Sollner2015, Vermersch2016, Lodahl2017, Albrecht2019, Jen2020_subradiance, Jen2020_PRR, Jen2020_disorder, Jen2021_bound, Zanner2022, Sheremet2022, Pennetta2022, Jen2022_correlation, Pennetta2022_2}. In this quantum interface, a strong coupling regime in light-matter interaction can be reached by coupling atoms with evanescent fields at the waveguide surface, where the directionality of light exchange processes can be further manipulated and tailored by external magnetic fields \cite{Mitsch2014}. This gives rise to exotic phenomena of collective radiation behaviors \cite{Lodahl2017, Albrecht2019, Jen2020_subradiance} and a new topological waveguide-QED platform that can host photonic bound states \cite{Kim2021}. 

\begin{figure}[t]
\centering
\includegraphics[width=0.5\textwidth]{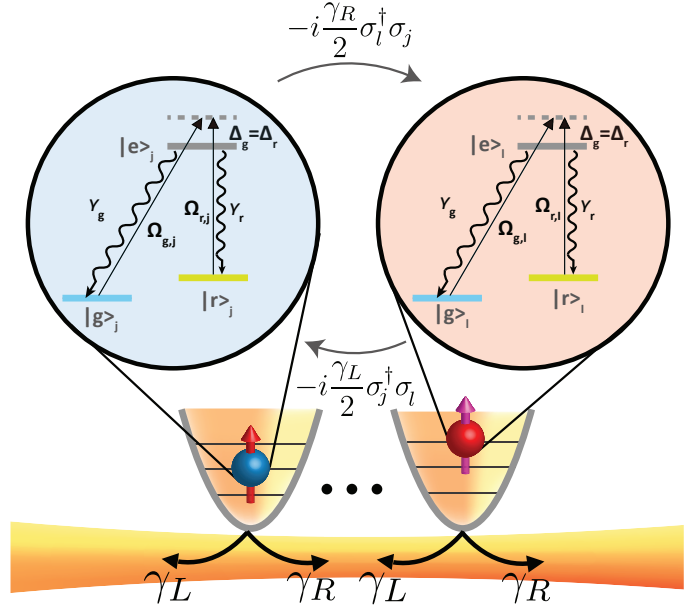}
\caption{A schematic plot of nonreciprocal couplings between trapped atoms. The $\Lambda$-type three-level atoms are trapped in harmonic potentials under the dark-state cooling scheme with two laser fields $\Omega_{g,j(l)}$ and $\Omega_{r,j(l)}$ operating on the transitions $|g\rangle_{j(l)}\rightarrow|e\rangle_{j(l)}$ and $|r\rangle_{j(l)}\rightarrow|e\rangle_{j(l)}$ respectively at two-photon resonance $\Delta_g=\Delta_r$. Respective decay rates for individual atoms are $\gamma_{g(r)}$ along with nonreciprocal decay channels $\gamma_L=\gamma_{g,L}+\gamma_{r,L}$ and $\gamma_R=\gamma_{g,R}+\gamma_{r,R}$ ($\gamma_{g(r)}$$=$$\gamma_{g(r),L}$$+$$\gamma_{g(r),R}$). These left (L)- and right (R)-propagating decay rates represent the nonreciprocal couplings $\gamma_{g(r),L}\neq\gamma_{g(r),R}$, which effectively facilitates spin-exchange hopping between the $j$th and $l$th sites of trapped atoms via an atom-waveguide interface.}\label{Fig1}
\end{figure}

In this work, we theoretically investigate the dark-state sideband cooling scheme in trapped atoms with nonreciprocal couplings, as shown in Fig. \ref{Fig1}. The nonreciprocal couplings can be facilitated in an atom-waveguide interface or via a photonic quantum link in free space \cite{Grankin2018}, which leads to collective spin-phonon correlations \cite{Jordan2019, Shankar2019} and distinct heat exchanges. Here we present a superior dark-state cooling compared to the conventional single atom results by utilizing the nonreciprocal couplings between atoms. We explore various tunable parameter regimes which are allowed in dark-state cooling and identify the parameter region that gives better cooling performance without compromising its cooling rate. We further obtain an analytical form of the phonon occupation for the target atom under the asymmetric driving conditions. Our results demonstrate a distinct heat transfer within the atoms, which leads to a superior dark-state sideband cooling assisted by collective spin-exchange interactions. This opens new avenues in surpassing the cooling obstacle \cite{Pino2021, Leibfried2003, Jordan2019}, which is crucial in scalable quantum computation \cite{Cirac1995, Kielpinski2002, Pino2021, Shen2020}, quantum simulations  \cite{Buluta2009, Lanyon2011}, and preparations of large ensemble of ultracold atoms \cite{Huang2021, Hsiao2018} or molecules \cite{Anderegg2018}. This paper is organized as follows. We present the theoretical model of dark-state cooling with nonreciprocal couplings in Sec. II. We then identify the parameter regimes that demonstrate superior cooling behaviors and their cooling dynamics in Sec. III. In Sec. IV, we further show the analytical prediction of the steady-state phonon occupation of the target atom and confirm its validity under dark-state mapping. Finally we conclude in Sec. V. 

%%%%%%%%%%%%%%%%%%%%%%%%%%%%%%%%%%%%%%%%%%%%%%%%%%%%%%%%%%%%%%
%%%%%%%%%%%%%%%%%%%%%%%%%%%%%%%%%%%%%%%%%%%%%%%%%%%%%%%%%%%%%

\section{Theoretical model} 

We start with a conventional EIT cooling scheme \cite{Morigi2000} with nonreciprocal couplings between constituent trapped atoms, as shown in Fig. \ref{Fig1}. Each atom involves $\Lambda$-type three-level atomic structure, where two hyperfine ground states $|g\rangle_j$ and $|r\rangle_j$ for $j$th atom couple to their common excited state $|e\rangle_j$ with laser fields of Rabi frequencies $\Omega_{g,j}$ and $\Omega_{r,j}$, respectively. The intrinsic decay rates from the excited state are $\gamma_g$ and $\gamma_r$. With the nonreciprocal couplings introduced as $\gamma_L$ and $\gamma_R$, the dynamics of a density matrix $\rho$ for $N$ atoms with mass $m$ can be expressed as ($\hbar$$=$$1$) 
\bea
\frac{d \rho}{dt}=-i[H_{\rm LD}+H_L+H_R,\rho]+\mathcal{L}_L[\rho]+\mathcal{L}_R[\rho],\label{rho}
\eea
where $H_{\rm LD}$ for the EIT cooling in the Lamb-Dicke (LD) regime (in the first order of LD parameter $\eta_{g(r)}$) reads
\bea
H_{\rm LD}=&&-\sum_{j=1}^N \Delta_j|e\rangle_j\langle e|+\nu\sum_{j=1}^N a_j^\dag a_j+\sum_{j=1}^N\left(\frac{\Omega_{g,j}}{2}|e\rangle_j\langle g|+\frac{\Omega_{r,j}}{2}|e\rangle_j\langle r|+\textrm{H.c.}\right)\nonumber\\
&&+i\eta_g\cos(\psi_g)\sum_{j=1}^N\left[\frac{\Omega_{g,j}}{2}|e\rangle_j\langle g|\left(a_j+a_j^\dag\right)-\textrm{H.c.}\right]+i\eta_r\cos(\psi_r)\sum_{j=1}^N\left[\frac{\Omega_{r,j}}{2}|e\rangle_j\langle r|\left(a_j+a_j^\dag\right)-\textrm{H.c.}\right],\label{LD}
\eea
with the common laser detuning $\Delta_j\equiv\omega_{g,j}-\omega_{eg}=\omega_{r,j}-\omega_{er}$ as required in EIT cooling scheme, i.e. the same difference between the $j$th-site laser central frequencies ($\omega_{g(r),j}$) and the atomic transition frequencies ($\omega_{eg}$ and $\omega_{er}$). Projection angles of the laser fields to the motional direction are denoted as $\psi_{g(r)}$. The harmonic trap frequency is $\nu$ with a creation (annihilation) operator $a_j^\dag$ ($a_j$) in the quantized phononic states $|n\rangle$, and LD parameters are $\eta_{g(r)}=k_{g(r)}/\sqrt{2m\nu}$ with $k_{g(r)}\equiv\omega_{g(r),j}/c$. The coherent and dissipative nonreciprocal couplings in the zeroth order of $\eta_{g,r}$ are \cite{Pichler2015} 
\bea
H_{L(R)} =&& -i\sum_{m=g,r}\frac{\gamma_{m,L(R)}}{2} \sum_{\mu<(>)\nu}^N\left(e^{ik_s|x_\mu-x_\nu|}|e\rangle_\mu\langle m|  \otimes|m\rangle_\nu\langle e|-\textrm{H.c.}\right),\label{LR}
\eea
and
\bea
\mathcal{L}_{L(R)}[\rho]=&&-\sum_{m=g,r}\frac{\gamma_{m,L(R)}}{2} \sum_{\mu,\nu=1}^N e^{\mp ik_s(x_\mu-x_\nu)} \left(|e\rangle_\mu\langle m|\otimes|m\rangle_\nu\langle e|\rho + \rho |e\rangle_\mu\langle m|\otimes|m\rangle_\nu\langle e|-2|m\rangle_\nu\langle e|\rho|e\rangle_\mu\langle m|\right),\label{Lindblad}
\eea
respectively, where $k_s$ denotes the wave vector in the guided mode that mediates nonreciprocal couplings $\gamma_{L(R)}\equiv\gamma_{g,L(R)}+\gamma_{r,L(R)}$, and $\xi\equiv k_s |x_{\mu+1}-x_{\mu}|$ quantifies the light-induced dipole-dipole interactions between the relative positions of trap centers $x_\mu$ and $x_\nu$. 

In terms of the eigenstates with only atomic spin degrees of freedom in Eq. (\ref{LD}), we can further diagonalize the EIT Hamiltonian \cite{Fleischhauer2005} in a single-particle limit, where three eigenstates are  
\bea
|+\rangle=\sin\phi|e\rangle-\cos\phi|b\rangle,~ |-\rangle=\cos\phi|e\rangle+\sin\phi|b\rangle,~ |d\rangle=\cos\theta|g\rangle-\sin\theta|r\rangle,
\eea  
with $|b\rangle\equiv\sin\theta|g\rangle+\cos\theta|r\rangle$, and the corresponding eigenenergies are 
\bea
\omega_+=\left(-\Delta+\sqrt{\Omega_g^2+\Omega_r^2+\Delta^2}\right)/2,~\omega_-=\left(-\Delta-\sqrt{\Omega_g^2+\Omega_r^2+\Delta^2}\right)/2,~\omega_d=0. 
\eea
The angles $\theta$ and $2\phi$ are defined as $\tan^{-1}(\Omega_g/\Omega_r)$ and $\tan^{-1}(-\sqrt{\Omega_g^2+\Omega_r^2}/\Delta)$, respectively. Under the EIT cooling condition of $\omega_+=\nu$ \cite{Morigi2000, Zhang2021}, a red sideband transition between $|d,n\rangle\leftrightarrow|+,n-1\rangle$ becomes resonant, and we can safely ignore the influences of off-resonant state $|-\rangle$ and blue sideband transition between $|d,n\rangle\leftrightarrow|+,n+1\rangle$. 

The effective Hamiltonian within the subspace $|d\rangle$ and $|+\rangle$ can then be obtained from Eq. (\ref{LD}), which reads in an interaction picture \cite{Zhang2021, Zhang2021_2},   
\bea
H_{\rm LD}^{\rm eff}=i\eta_{\rm eff}\sum_{j=1}^N\frac{\Omega_{{\rm eff}, j}}{2}\left[|d\rangle_j\langle +|\left(a_j^\dag+a_j\right)-\textrm{H.c.}\right],\label{effLD}
\eea
with $\eta_{\rm eff}\equiv\eta_g\cos(\psi_g)-\eta_r\cos(\psi_r)$, $\Omega_{{\rm eff}, j}\equiv\Omega_{g,j}\Omega_{r,j}\sin\phi_j/\sqrt{\Omega_{g,j}^2+\Omega_{r,j}^2}$, $\tan(2\phi_j)\equiv-\sqrt{\Omega_{g,j}^2+\Omega_{r,j}^2}/\Delta_j$, and $\Delta_j=-\nu+(\Omega_{g,j}^2+\Omega_{r,j}^2)/(4\nu)$. The associated nonreciprocal coupling terms as in Eqs. (\ref{LR}) and (\ref{Lindblad}) reduce to 
\bea
H_{L(R)}^{\rm eff}=&&-i \sum_{\mu<(>)\nu}^N\frac{\gamma_{{\rm eff},L(R)}(\mu,\nu)}{2}\left(e^{ik_s|x_\mu-x_\nu|}|+\rangle_\mu\langle d|  \otimes|d\rangle_\nu\langle +|-\textrm{H.c.}\right),\label{effLR}\\
\mathcal{L}_{L(R)}^{\rm eff}[\rho]=&&-\sum_{\mu\neq\nu,\nu=1}^N \frac{\gamma_{{\rm eff},L(R)}(\mu,\nu)}{2} e^{\mp ik_s(x_\mu-x_\nu)} \left(|+\rangle_\mu\langle d|\otimes|d\rangle_\nu\langle +|\rho + \rho |+\rangle_\mu\langle d|\otimes|d\rangle_\nu\langle +|-2|d\rangle_\nu\langle +|\rho|+\rangle_\mu\langle d|\right)\nonumber\\
&&-\sum_{\mu=1}^N \frac{\gamma_{{\rm eff},\mu}}{2}\left(|+\rangle_\mu\langle +|\rho + \rho |+\rangle_\mu\langle +|-2|d\rangle_\mu\langle +|\rho|+\rangle_\mu\langle d|\right)
,\label{effLindblad}
\eea 
where  
\bea
\gamma_{{\rm eff},L(R)}(\mu,\nu)\equiv&& \sin\phi_\mu \sin\phi_\nu\left[\gamma_{g,L(R)} \cos\theta_\mu\cos\theta_\nu + \gamma_{r,L(R)}\sin\theta_\mu\sin\theta_\nu\right], \\ 
\gamma_{{\rm eff},\mu}\equiv&& \sin^2\phi_\mu\left(\gamma_g \cos^2\theta_\mu + \gamma_r \sin^2\theta_\mu \right),
\eea
with $\gamma_{g(r)}$$=$$\gamma_{g(r),L}$$+$$\gamma_{g(r),R}$ and $\tan\theta_\mu= \Omega_{g,\mu}/\Omega_{r,\mu}$. We note that $\gamma_{{\rm eff}, \mu}\gamma_{{\rm eff}, \nu}\neq[\gamma_{{\rm eff},L}(\mu,\nu)+\gamma_{{\rm eff},R}(\mu,\nu)]^2$ in general, unless $\theta_\mu=\theta_\nu$, which would otherwise be assured in quantum systems with collective and pairwise dipole-dipole interactions \cite{Pichler2015}. This inequality arises owing to the mapping to the reduced Hilbert space, where different $\phi_\mu$ and $\theta_\mu$ determined by external laser fields can further allow extra and tunable degrees of freedom for cooling mechanism.  

The above dark-state mapping sustains when the state $|-\rangle$ can be detuned significantly from the subspace $|+,n-1\rangle$ and $|d,n\rangle$, which leads to the requirement of $\Delta_j\gg\Omega_{r,j}\gg\nu$ for the $j$th atom based on the condition of $|\omega_-|\gg\nu$, the EIT cooling condition $\omega_+=\nu$, and the usual assumption of a weak probe field $\Omega_{g,j}\ll\Omega_{r,j}$. The resultant effective dark-state sideband cooling should also be legitimate by fulfilling the sideband cooling condition of $\nu\gg\eta_{\rm eff}\Omega_{{\rm eff},j}$ and $\gamma_{{\rm eff},j}$. This can be guaranteed when $\phi_j$ and $\theta_j$ are made small enough, which can easily be achieved by choosing large enough control fields $\Omega_{r,j}$ and so are $\Delta_j$. In the setting of EIT cooling with nonreciprocal couplings, it is these tunable excitation parameters and the controllable directionality of spin-exchange couplings that give rise to distinct cooling behaviors we explore in the following section.   

\section{Superior dark-state cooling}

Here we consider a setting of two atoms with $\Lambda$-type three-level configurations with nonreciprocal couplings, which represents the building block for a large-scale atomic array. In this basic unit under EIT cooling with collective spin-exchange interactions, we numerically obtain the steady-state phonon occupations tr$[\rho(t\rightarrow\infty) a^\dag_{j}a_j]=\langle n_{j}\rangle_{\rm ss}$ from Eq. (\ref{rho}), under the Hilbert space $|\alpha, n\rangle_j$ with $\alpha\in\{g,r,e\}$ and $n\in\{0,1,2\}$. This truncation of phonon numbers is valid when the composite two-atom system is close to their motional ground state, that is when $\langle n_{j}\rangle_{\rm ss}\ll 1$.

\begin{figure}[t]
\centering
\includegraphics[width=0.95\textwidth]{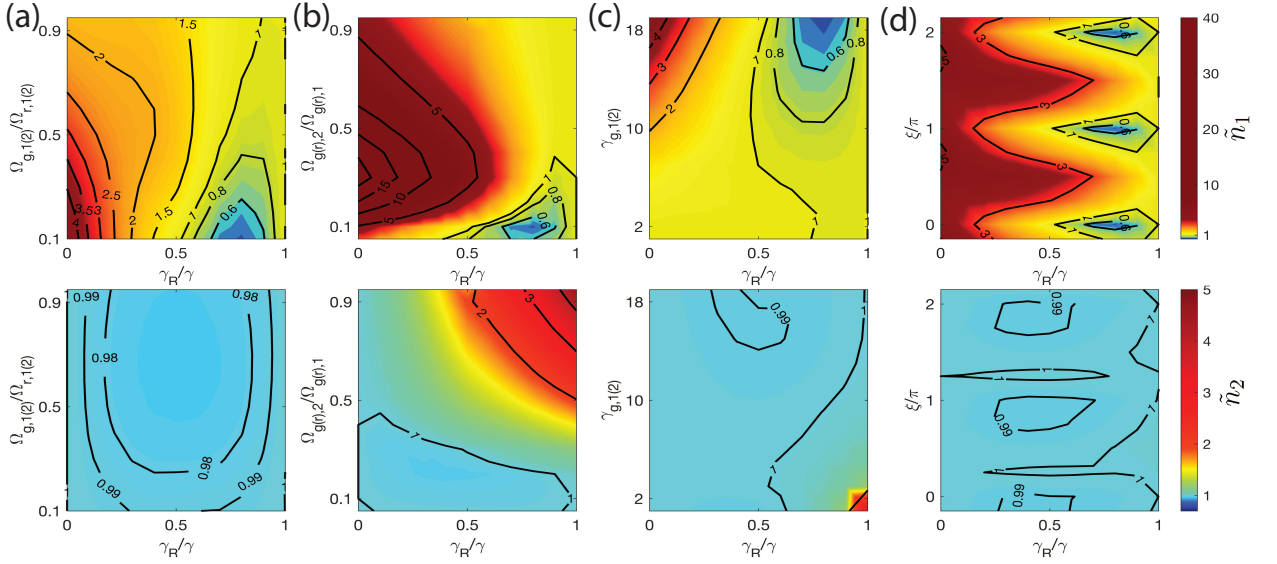}
\caption{Superior cooling regions for the target atom ($\tilde{n}_{1}$) in a two-atom system under an asymmetrically-driven EIT cooling. The upper and lower panels are plotted for $\tilde{n}_{1}$ and $\tilde{n}_{2}$ respectively comparing to their single atom results. We explore the cooling parameter regimes for (a) $\Omega_{g,j}/\Omega_{r,j}$ with $\Omega_{g(r),2}/\Omega_{g(r),1}=0.1$, (b) $\Omega_{g(r),2}/\Omega_{g(r),1}$ with $\Omega_{g,j}/\Omega_{r,j}=0.1$, (c) $\gamma_{g,j}/\nu$ with $\gamma_{g,j}+\gamma_{r,j}=20\nu$, and (d) $\xi/\pi$. Other common parameters are chosen as $\Omega_{g(r),2}/\Omega_{g(r),1}=0.1$, $\Omega_{g,j}/\Omega_{r,j}=0.1$, $\Omega_{r,1}=15\nu$, $\eta_{g(r)}=0.15$, $\gamma_{g,j}=18\nu$, $\gamma_{r,j}=2\nu$, $\xi/\pi=2$, $\psi_{g,j}=\pi/4$, and $\psi_{r,j}=3\pi/4$. A normalization of decay rate is $\gamma\equiv\gamma_R+\gamma_L$.}\label{Fig2}
\end{figure}

As a comparison to the single atom result ($\langle n_j\rangle_{\rm ss}^{\rm s}$) without nonreciprocal spin exchange interactions, we calculate the ratio $\tilde n_j\equiv\langle n_{j}\rangle_{\rm ss}/\langle n_j\rangle_{\rm ss}^{\rm s}$ to account for the superior or inferior cooling regime when $\tilde n_j$ is less or more than $1$. In Fig. \ref{Fig2}, we explore various parameter regimes of $\Omega_{g,j}$, $\Omega_{r,j}$, $\gamma_{g,j}$, $\xi$, $\gamma_R$ and look for superior cooling regions in the allowable parameters under EIT cooling conditions. In all plots, we investigate an asymmetric setup of driving fields in two atoms, where we focus on the cooling behavior of the target atom ($\tilde n_1$), or otherwise we always observe heating effect. In Fig. \ref{Fig2}(a), we find a superior cooling regime for the target atom when $\gamma_{R}/\gamma>0.5$ with a minimum at $\gamma_R/\gamma\approx 0.75$ for a smaller probe-to-control field ratio $\Omega_{g,j}/\Omega_{r,j}$. This presents the essential effect of nonreciprocal couplings, which leads to surpassing cooling behaviors owing to distinct heat transfer from collective spin-exchange interactions. The heating effect from symmetric EIT settings can be seen in Fig. \ref{Fig2}(b), where $\tilde n_{1(2)}$ at $\gamma_R/\gamma<(>)0.5$ becomes larger than the respective single atom cases when $\Omega_{g(r),2}/\Omega_{g(r),1}=1$. We note that at the unidirectional coupling when $\gamma_R=1(0)$, the $\tilde n_{1(2)}$ reaches the single atom limit as if the atom experiences one-way decay channel without the back action from the other atom. The asymmetric setting is favored for superior cooling and evidenced in Fig. \ref{Fig2}(b) when $\Omega_{g(r),2}/\Omega_{g(r),1}<0.5$. 

%%%%%%%%%%%%%%%%%%%%%%%%%%%%%%%%%%%%%%%%%%%%%%%%%%%%%%%%%%%%%%
%%%%%%%%%%%%%%%%%%%%%%%%%%%%%%%%%%%%%%%%%%%%%%%%%%%%%%%%%%%%%

Furthermore in Fig. \ref{Fig2}(c), we explore the effect of decay channels in the probe field and control field transitions. As $\gamma_g$ increases along with a finite nonreciprocal coupling $\gamma_R$, a superior cooling region emerges and allows an impressive performance of almost twofold improvement ($\tilde n_1\approx 0.5$) compared to the case without spin-exchange interactions. Finally in Fig. \ref{Fig2}(d), we find the optimal operations of EIT cooling at $\xi=2\pi$ or $\pi$, which also reflects the optimal interparticle distances that permit superior EIT cooling. For $\tilde n_2$ of the residual atom in all lower panels, we find no significant cooling or even heating behaviors under asymmetric EIT driving conditions. This shows the essential role of the residual atom that hosts spin-exchange couplings to remove extra heat from the target atom. We note that throughout the paper, we consider the same ratio of nonreciprocal couplings between the left- and the right-propagating channels in either $\gamma_g$ or $\gamma_r$ for simplicity. 

\begin{figure}[t]
\centering
\includegraphics[width=0.95\textwidth]{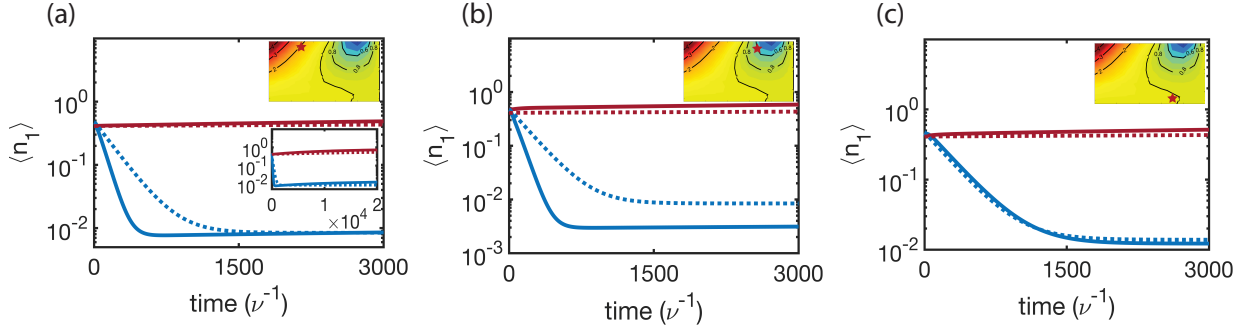}
\caption{Time dynamics of average phonon occupations for the target atom. We numerically simulate the time dynamics of $\langle n_1\rangle$ (blue-solid line) and $\langle n_2\rangle$ (red-solid line) with specific parameters denoted by star polygons in the inset plots extracted from Fig. \ref{Fig2}(c), for (a) $(\gamma_g/\nu,\gamma_R/\gamma)=(18,0.3)$, (b) $(18,0.7)$, and (c) $(2,0.7)$, in heating, cooling, and neutral regimes, respectively. The respective dotted lines represent the corresponding results of a single atom under the EIT cooling scheme without nonreciprocal couplings. All $\langle n_2\rangle$ almost overlap with respective single atom results, which signifies no significant cooling or heating effect as evidenced in the steady-state solutions in Fig. \ref{Fig2}(c). The lower inset plot in (a) shows long-time behaviors of the heating region approaching the steady states. The numerical simulation uses $n_0=0.7$ and truncates the phonon states to $n=2$.}\label{Fig3}
\end{figure}

Next we numerically simulate the time dynamics of two-atom system in Fig. \ref{Fig3}. To genuinely simulate the time evolutions in Eq. (\ref{rho}), we postulate the system in a thermal state initially \cite{Zhang2021, Roos2000_2}, 
\bea
\rho(t=0)=\prod_{\mu=1}^N \sum_{n=0}^\infty \frac{n_0^n}{(n_0+1)^{n+1}} |g, n\rangle_\mu\langle g, n|,
\eea
where $n_0$ is an average phonon number for both atoms. We assume $n_0\lesssim 1$ and make a sufficient truncation on the finite motional states to guarantee the convergence in numerical simulations. As shown in Fig. \ref{Fig3}, we select three contrasted regimes for heating, cooling, and neutral time dynamics. The characteristic times to reach the steady-states are shown shorter in cooling dynamics in Figs. \ref{Fig3}(b). This presents an enhancement in cooling rate compared to the EIT cooling scheme in a single atom, in addition to the outperforming cooling behaviors in the steady-state phonon occupations. By contrast, the time dynamics in the heating regime shows a longer time behavior than the single atom case, while the neutral regime has a similar characteristic timescale. We attribute the enhanced cooling rate or prolonged time dynamics in heating to the collective spin-phonon couplings between the atoms.

%%%%%%%%%%%%%%%%%%%%%%%%%%%%%%%%%%%%%%%%%%%%%%%%%%%%%%%%%%%%%%
%%%%%%%%%%%%%%%%%%%%%%%%%%%%%%%%%%%%%%%%%%%%%%%%%%%%%%%%%%%%%

\section{Dark-state sideband cooling under asymmetric drivings}

In Eqs. (\ref{effLD}-\ref{effLindblad}), we have demonstrated the mapping from EIT cooling with three-level structures to the effective dark-state sideband cooling scheme. Within $n\in\{0,1\}$, the mapping has reduced the Hilbert space dimensions in two atoms from $36$ to $16$ with $16^2$ coupled linear equations. We follow the wisdom in sideband cooling with chiral couplings by considering asymmetric driving conditions when $\eta_{\rm eff}\Omega_{{\rm eff},2}\ll\eta_{\rm eff}\Omega_{{\rm eff},1}$ \cite{Chen2022}. In this way, we are able to perform a partial trace on the motional degree of freedom ($a_{2}$) and focus on the target atom behavior $\langle n_1\rangle_{\rm ss}$, which further diminishes the dimension of the Hilbert space to $8$. The dynamics of this two-atom system can then be determined by the reduced density matrix $\text{tr}_{a_{2}}(\rho)$. 

The steady-state solutions for the density matrix elements of $\rho_{\mu_{1}n_{1}\mu_{2}\nu_{1}m_{1}\nu_{2}}=\langle\mu_{1},n_{1};\mu_{2}|\text{tr}_{a_{2}}\rho|\nu_{1},m_{1};\nu_{2}\rangle$ can further be simplified by taking advantage of the fact that $\rho_{g0gg0g}$ is $\mathcal{O}(1)$ when $\gamma_{\rm eff}^2/\nu^2$ and $\eta_{\rm eff}^2\Omega_{{\rm eff},j}^2/\nu^2$ are much smaller than one. Therefore, we can neglect those $\rho_{\mu_{1}n_{1}\mu_{2}\nu_{1}m_{1}\nu_{2}}$ whose leading terms are higher than the second order. Setting \(e^{ikd}=1\) and keeping the terms of $\mathcal{O}(1)$, $\mathcal{O}(\gamma_{{\rm eff},1(2)})$, $\mathcal{O}(\gamma_{{\rm eff},L(R)})$, $\mathcal{O}(\eta_{\rm eff}\Omega_{{\rm eff},1})$, $\mathcal{O}(\gamma_{{\rm eff},1(2,L,R)}\eta_{\rm eff}\Omega_{{\rm eff},1})$, we obtain 	
	\begin{equation}
	\begin{split}
		0 = & \eta_{\rm eff}\Omega_{{\rm eff},1}(\rho_{e0gg1g}-\rho_{g1ge0g})+2i\gamma_{{\rm eff},1}\rho_{e0ge0g}+2i\gamma_{{\rm eff},L}(1,2)(\rho_{g0ee0g}+\rho_{e0gg0e}),\\
		0 = & -\eta_{\rm eff}\Omega_{{\rm eff},1}\rho_{g0eg1g}-2i\gamma_{{\rm eff},L}(1,2)\rho_{g0eg0e}-2i\gamma_{{\rm eff},R}(1,2)\rho_{e0ge0g}-i(\gamma_{{\rm eff},1}+\gamma_{{\rm eff},2})\rho_{g0ee0g},\\
		0 = & \eta_{\rm eff}\Omega_{{\rm eff},1}(\rho_{e0ge0g}-\rho_{g1gg1g})-2i\gamma_{{\rm eff},L}(1,2)\rho_{g1gg0e}-i\gamma_{{\rm eff},1}\rho_{g1ge0g},\\
		0 = & -2i\gamma_{{\rm eff},R}(1,2)(\rho_{e0gg0e}+\rho_{g0ee0g})-2i\gamma_{{\rm eff},2}\rho_{g0eg0e},\\
		0 = & \eta_{\rm eff}\Omega_{{\rm eff},1}\rho_{e0gg0e}-2i\gamma_{{\rm eff},R}(1,2)\rho_{g1ge0g}-i\gamma_{{\rm eff},2}\rho_{g1gg0e},\\
		0 = & \eta_{\rm eff}\Omega_{{\rm eff},1}(\rho_{e0gg1g}-\rho_{g1ge0g})+i\gamma_{{\rm eff},1}\frac{\eta_{\rm eff}^{2}\Omega_{{\rm eff},1}^{2}}{8\nu^{2}}\rho_{g0gg0g},
		\end{split}
	\end{equation}
along with three other relationships, 
	\begin{align}\label{relation}
	\begin{split}
		\rho_{e1ge1g}  =  \frac{\eta_{\rm eff}^{2}\Omega_{{\rm eff},1}^{2}}{16\nu^{2}}\rho_{g0gg0g},~
		\rho_{g1eg0g}   =  -\frac{i\gamma_{{\rm eff},R}(1,2)\eta_{\rm eff}\Omega_{{\rm eff},1}}{8\nu^{2}}\rho_{g0gg0g},~
		\rho_{e1gg0g}   =  -\frac{4\nu-i\gamma_{{\rm eff},2}}{16\nu^{2}}\eta_{\rm eff}\Omega_{{\rm eff},1}\rho_{g0gg0g}.
		\end{split}
	\end{align}
Solving the above equations leads to the solutions of density matrix elements expressed in terms of \(\rho_{g0gg0g}\). Finally we obtain the steady-state phonon occupation for the target atom as (\(\rho_{g0gg0g}\approx 1\))
	\begin{eqnarray}
	\langle n_{1}\rangle_{\text{st}} = && \rho_{e1ee1e}+\rho_{e1ge1g}+\rho_{g1eg1e}+\rho_{g1gg1g},\nonumber \\
	 \approx && \frac{\gamma_{{\rm eff},1}^2}{16\nu^{2}}+\frac{\eta_{\rm eff}^{2}\Omega_{{\rm eff},1}^{2}}{8\nu^{2}}-\frac{\gamma_{{\rm eff},1}}{\gamma_{{\rm eff},2}}\frac{\gamma_{{\rm eff},R}(1,2)\gamma_{{\rm eff},L}(1,2)}{4\nu^{2}}\nonumber\\
	&&+\frac{(\gamma_{{\rm eff},1}+\gamma_{{\rm eff},2})^2/(\gamma_{{\rm eff},1}\gamma_{{\rm eff},2})[\gamma_{{\rm eff},R}(1,2)\gamma_{{\rm eff},L}(1,2)]}{(\gamma_{{\rm eff},1}+\gamma_{{\rm eff},2})/\gamma_{{\rm eff},1}\left[\gamma_{{\rm eff},1}\gamma_{{\rm eff},2}/4-\gamma_{{\rm eff},R}(1,2)\gamma_{{\rm eff},L}(1,2)\right]+\eta_{\rm eff}^{2}\Omega_{{\rm eff},1}^{2}/4}\frac{\eta_{\rm eff}^{2}\Omega_{{\rm eff},1}^{2}}{16\nu^{2}},\label{n1}
	\end{eqnarray}	
where the approximate form assumes $\nu\gg\gamma_{{\rm eff},L(R)}(1,2),\gamma_{{\rm eff}, 1(2)}$, and the first two terms are exactly the steady-state phonon occupation for a single atom under the sideband cooling \cite{Zhang2021}. The remaining terms are the modifications originating from the nonreciprocal couplings. At unidirectional coupling $\gamma_{{\rm eff},L(R)}(1,2)=0$, Eq. (\ref{n1}) again reduces to the single atom result, which is also shown in Figs. \ref{Fig2}(a) and \ref{Fig2}(b) when $\gamma_R/\gamma=1$ or $0$.  
	
%%%%%%%%%%%%%%%%%%%%%%%%%%%%%%%%%%%%%%%%%%%%%%%%%%%%%%%%%%%%%%
\begin{figure}[t]
\centering
\includegraphics[width=0.6\textwidth]{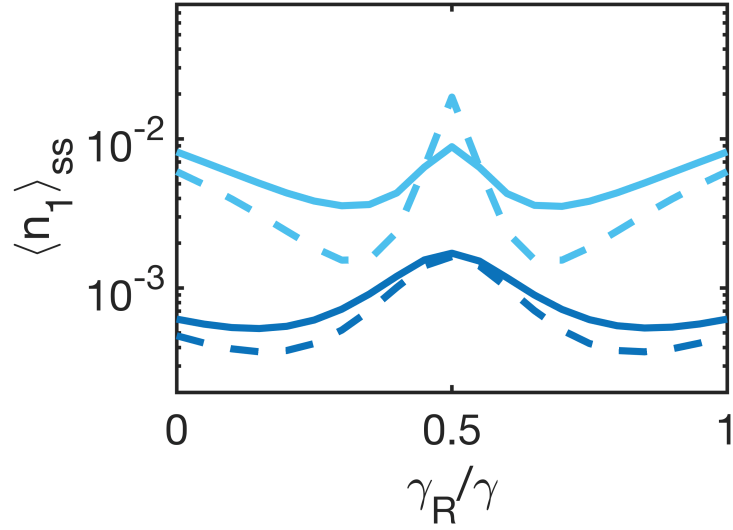}
\caption{Steady-state occupations of two-atom system under dark-state sideband cooling. Under the asymmetric driving condition in the dark-state sideband cooling scheme, the motional degrees of freedom of the residual atom $a_2$ can be traced out and Eq. (\ref{n1}) can be obtained analytically. For the target atom, we numerically calculate $\langle n_1\rangle_{\rm ss}$ from the EIT cooling scheme with nonreciprocal couplings with three-level atomic structures (solid line) and compare it with Eq. (\ref{n1}) from the effective dark-state sideband cooling scheme under the dark-state mapping (dashed line). The driving fields in the two-atom EIT cooling with a light blue to a darker blue line are chosen as $(\Omega_{g,1},\Omega_{r,1},\Omega_{g,2},\Omega_{r,2})/\nu=(1.5, 15, 0.015, 15)$ and $(3, 30, 0.03, 30)$, respectively, with corresponding larger two-photon detunings $\Delta_{1(2)}$ in the latter case. We choose $\eta_{g(r)}=0.1$, $\gamma_{g}=18\nu$, $\gamma_{r}=2\nu$, and the other relevant parameters are the same as the common parameters we apply in Fig. \ref{Fig2}.}\label{Fig4}
\end{figure}
%%%%%%%%%%%%%%%%%%%%%%%%%%%%%%%%%%%%%%%%%%%%%%%%%%%%%%%%%%%%%

In the LD regime as $\eta_{\rm eff}\rightarrow 0$, we can estimate Eq. (\ref{n1}) as 
\bea
\langle n_{1}\rangle_{\text{ss}}|_{\eta_{\rm eff}\rightarrow 0} \approx \frac{\gamma_{{\rm eff},1}^2}{16\nu^{2}}-\frac{\gamma_{{\rm eff},1}}{\gamma_{{\rm eff},2}}\frac{\gamma_{{\rm eff},R}(1,2)\gamma_{{\rm eff},L}(1,2)}{4\nu^{2}},
\eea
where the phonon occupation of the target atom is no longer limited by the effective spontaneous emission rate as in the single atom result of $\gamma_{{\rm eff},1}^2/(16\nu^{2})$. If we assume $\theta_1=\theta_2\approx 0$ and $\gamma_g\gg\gamma_r$, by comparing to the single atom result, we obtain  
\bea
\tilde n_1|_{\eta_{\rm eff}\rightarrow 0} \approx 1-\frac{4\gamma_{g,L}\gamma_{g,R}}{\gamma_g^2}.
\eea
The above shows the minimum of $\tilde n_1|_{\eta_{\rm eff}\rightarrow 0}=0$ when $\gamma_{g,L(R)}=1/2$. This coincides with the predictions of the target ion under chiral-coupling-assisted sideband cooling \cite{Chen2022} when an extreme LD regime as $\eta\rightarrow 0$ is applied. Furthermore in this extreme regime as in an infinite mass limit, a reciprocal coupling emerges to host the heat removal in the target atom from the mechanism of collective spin-exchange interaction. The values of $\gamma_{g,R}$ at the minimum of $\tilde n_1$ would shift to nonreciprocal coupling regimes of $\gamma_{g,R}\lessgtr 0.5$ when finite corrections of $\eta_{\rm eff}$ in Eq. (\ref{n1}) are retrieved, where two minimums arise from a quartic dependence of $\gamma_{g,R}$ in Eq. (\ref{n1}). In Fig. \ref{Fig4}, we show the superior dark-state sideband cooling under nonreciprocal couplings, where two lowest minimums can be identified as superior cooling compared to the single atom results at $\gamma_R=1$. We compare the exact results from EIT cooling scheme with three-level atomic structures and the predictions from the effective dark-state sideband cooling, which are well described by an analytical form in Eq. (\ref{n1}) under asymmetric driving conditions. This confirms the validity of dark-state mapping when large enough two-photon detunings $\Delta_j$ are applied. 

\section{Conclusion}

In summary, we have shown theoretically that a superior EIT cooling behavior can be feasible when nonreciprocal couplings between constituent atoms are introduced. This provides a way to get around the cooling bottleneck \cite{Leibfried2003, Jordan2019, Pino2021} that an atom-based quantum computer or other applications of quantum technology may encounter. The conventional EIT cooling in a single atom is limited by the steady-state phonon occupation $\sim(\gamma/\Delta)^2$ \cite{Zhang2021_2}, while the corresponding dark-state sideband cooling is by $\sim(\gamma_{\rm eff}/\nu)^2$. By utilizing nonreciprocal couplings between constituent atoms, an intriguing dark-state cooling scheme in $\Lambda$-type three-level structure manifests a distinct heat transfer to further cool the target atom surpassing its single atom limit. We explore the allowable system parameters especially under asymmetric driving conditions and identify the regions for superior cooling performance with an enhanced cooling rate. Our results unravel the mechanism of collective spin-exchange interactions between trapped atoms, which enable new possibilities in driving the atoms toward their motional ground states. Future work will consider a multiatom platform, where more complex spin-phonon correlations may arise to further enhance the performance of the EIT cooling scheme. 

%%%%%%%%%%%%%%%%%%%%%%%%%%%%%%%%%%%%%%%%%%%%%%%%%%%%%%%%%%%%%%%%%%%%%%%%%%%%%%
\section*{ACKNOWLEDGMENTS}
We acknowledge support from the Ministry of Science and Technology (MOST), Taiwan, under Grant No. MOST-109-2112-M-001-035-MY3. We are also grateful for support from TG 1.2 and TG 3.2 of NCTS. 
%\appendix
%\section{Analytical form of the steady-state occupation of the target ion}

%%%%%%%%%%%%%%%%%%%%%%%%%%%%%%%%%%%%%%%%%%%%

\begin{thebibliography}{99}
%%%%%%%%%%%%%%%%%%%%%%%%%%%%%%%%%%% intro: book
\bibitem{Cohen2011} C. Cohen-Tannoudji and D. Gu{\'e}ry-Odelin, {\it Advances in Atomic Physics: An Overview} (World Scientific, 2011).
%%%%%%%%%%%%%%%%%%%%%%%%%%%%%%%%%%% intro:ions sideband cooling
\bibitem{Diedrich1989} F. Diedrich, J. C. Bergquist, W. M. Itano, and D. J. Wineland, Phys. Rev. Lett. {\bf 62}, 403 (1989). 
%%%%%%%%%%%%%%%% sideband cooling formula
\bibitem{Cirac1992} J. I. Cirac, R. Blatt, P. Zoller, and W. D. Phillips, Phys. Rev. A {\bf 46}, 2668 (1992).
%%%%%%%%%%%%%%%%%%%%%%%%%%%%%%%%%%% intro:ions sideband cooling
\bibitem{Monroe1995} C. Monroe, D. M. Meekhof, B. E. King, S. R. Jefferts, W. M. Itano, D. J. Wineland, and P. Gould, Phys. Rev. Lett. {\bf 75}, 4011 (1995).
\bibitem{Roos1999} C. Roos, T. Zeiger, H. Rohde, H. C. N\"{a}gerl, J. Eschner, D. Leibfried, F. Schmidt-Kaler, and R. Blatt, Phys. Rev. Lett. {\bf 83}, 4713 (1999).
\bibitem{Zhang2021} S. Zhang, J.-Q. Zhang, W. Wu, W.-S. Bao, and C. Guo, New J. Phys. {\bf 23}, 023018 (2021). 
%%%%%%%%%%%%%%%%%%%%%%%%%%%%%%%%%%% intro: EIT cooling
\bibitem{Morigi2000} G. Morigi, J. Eschner, and C. H. Keitel, Phys. Rev. Lett. {\bf 85}, 4458 (2000). 
\bibitem{Roos2000} C. F. Roos, D. Leibfried, A. Mundt, F. Schmidt-Kaler, J. Eschner, and R. Blatt, Phys. Rev. Lett. {\bf 85}, 5547 (2000).
\bibitem{Morigi2003} G. Morigi, Phys. Rev. A {\bf 67}, 033402 (2003). 
\bibitem{Retzker2007} A. Retzker and M. B. Plenio, New. J. Phys. {\bf 9}, 279 (2007).
\bibitem{Cerrillo2010} J. Cerrillo, A. Retzker, and M. B. Plenio, Phys. Rev. Lett. {\bf 104},
043003 (2010).
\bibitem{Albrecht2011} A. Albrecht, A. Retzker, C. Wunderlich, and M. B. Plenio, New J. Phys. {\bf 13}, 033009 (2011).
\bibitem{Zhang2012} S. Zhang, C.-W. Wu, and P.-X. Chen, Phys. Rev. A {\bf 85}, 053420 (2012).
\bibitem{Zhang2014} S. Zhang, Q.-H. Duan, C. Guo, C.-W. Wu, W. Wu, and P.-X. Chen, Phys. Rev. A {\bf 89}, 013402 (2014). 
\bibitem{Kampschulte2014} T. Kampschulte, W. Alt, S. Manz, M. Martinez-Dorantes, R. Reimann, S. Yoon, D. Meschede, M. Bienert, and G. Morigi, Phys. Rev. A {\bf 89}, 033404 (2014).
\bibitem{Lechner2016} R. Lechner, C. Maier, C. Hempel, P. Jurcevic, B. P. Lanyon, T. Monz, M. Brownnutt, R. Blatt, and C. F. Roos, Phys. Rev. A {\bf 93}, 053401 (2016). 
\bibitem{Scharnhorst2018} N. Scharnhorst, J. Cerrillo, J. Kramer, I. D. Leroux, J. B. Wübbena, A. Retzker, and P. O. Schmidt, Phys. Rev. A {\bf 98}, 023424 (2018).
\bibitem{Jordan2019} E. Jordan, K. A. Gilmore, A. Shankar, A. Safavi-Naini, J. G. Bohnet, M. J. Holland, and J. J. Bollinger, Phys. Rev. Lett. {\bf 122}, 053603 (2019).
\bibitem{Feng2020} L. Feng, W. L. Tan, A. De, A. Menon, A. Chu, G. Pagano, and C. Monroe, Phys. Rev. Lett. {\bf 125}, 053001 (2020). 
\bibitem{Qiao2021} M. Qiao, Y. Wang, Z. Cai, B. Du, P. Wang, C. Luan, W. Chen, H.-R. Noh, and K. Kim, Phys. Rev. Lett. {\bf 126}, 023604 (2021). 
\bibitem{Huang2021} C. Huang, S. Chai, and S.-Y. Lan, Phys. Rev. A {\bf 103}, 013305 (2021).
\bibitem{Zhang2021_2} S. Zhang, T.-C. Tian, Z.-Y. Wu, Z.-S. Zhang, X.-H. Wang, W. Wu, W.-S. Bao, and C. Guo, Phys. Rev. A {\bf 104}, 013117 (2021).
%%%%%%%%%%%%%%%% optomechanical cooling
\bibitem{Xu2019} H. Xu, L. Jiang, A. A. Clerk, and J. G. E. Harris, Nature {568}, 65 (2019). 
\bibitem{Lai2020} D.-G. Lai, J.-F. Huang, X.-L. Yin, B.-P. Hou, W. Li, D. Vitali, F. Nori, and J.-Q. Liao, Phys. Rev. A {\bf 102}, 011502(R) (2020).
%%%%%%%%%%%%%%%%%%%%%%%%%%%%%%%%%%% intro: infinite couplings and review
\bibitem{Chang2018} D. E. Chang, J. S. Douglas, A. Gonz\'alez-Tudela, C.-L. Hung, H. J. Kimble, Rev. Mod. Phys. {\bf 90}, 031002 (2018).
\bibitem{Corzo2019} N. V. Corzo, J. Raskop, A. Chandra, A. S. Sheremet, B. Gouraud, and J. Laurat, Nature {\bf 566}, 359 (2019).
\bibitem{Sheremet2022} A. S. Sheremet, M. I. Petrov, I. V. Iorsh, A. V. Poshakinskiy, A. N. Poddubny, arXiv:2103.06824.
%%%%%%%%%%%%%%%%%%%%%%%%%%%%%%%%%%% intro: chiral
%%%%%%%%%%%%%%%% cascaded
\bibitem{Gardiner1993} C. W. Gardiner, Phys. Rev. Lett. {\bf 70} 2269 (1993). 
\bibitem{Carmichael1993} H. J. Carmichael, Phys. Rev. Lett. {\bf 70} 2273 (1993).
\bibitem{Stannigel2012} K. Stannigel, P. Rabl, and P. Zoller, New J. Phys. {\bf 14}, 063014 (2012).
\bibitem{Luxmoore2013} I. J. Luxmoore, N. A. Wasley, A. J. Ramsay, A. C. T. Thijssen, R. Oulton, M. Hugues, S. Kasture, V. G. Achanta, A. M. Fox, and M. S. Skolnick, Phys. Rev. Lett. {\bf 110}, 037402 (2013).
\bibitem{Ramos2014} T. Ramos, H. Pichler, A. J. Daley, and P. Zoller, Phys. Rev. Lett. {\bf 113}, 237203 (2014).
\bibitem{Arcari2014} M. Arcari, I. S\"{o}llner, A. Javadi, S. Lindskov Hansen, S. Mahmoodian, J. Liu, H. Thyrrestrup, E. H. Lee, J. D. Song, S. Stobbe, and P. Lodahl, Phys. Rev. Lett. {\bf 113}, 093603 (2014).
\bibitem{Mitsch2014} R. Mitsch, C. Sayrin, B. Albrecht, P. Schneeweiss, and A. Rauschenbeutel, Nat. Commun. {\bf 5}, 5713 (2014).
\bibitem{Pichler2015} H. Pichler, T. Ramos, A. J. Daley, and P. Zoller, Phys. Rev. A {\bf 91}, 042116 (2015).
\bibitem{Sollner2015} I. S\"ollner, S. Mahmoodian, S. L. Hansen, L. Midolo, A. Javadi, G. Kiršanskė, T. Pregnolato, H. El-Ella, E. H. Lee, J. D. Song, {\it et al.}, Nat. Nanotechnol. {\bf 10}, 775 (2015).
\bibitem{Vermersch2016} B. Vermersch, T. Ramos, P. Hauke, and P. Zoller, Phys. Rev. A {\bf 93}, 063830 (2016). 
\bibitem{Lodahl2017} P. Lodahl, S. Mahmoodian, S. Stobbe, A. Rauschenbeutel, P. Schneeweiss, J. Volz, H. Pichler, and P. Zoller, Nature {\bf 541}, 473 (2017).
\bibitem{Albrecht2019} A. Albrecht, L. Henriet, A. Asenjo-Garcia, P. B Dieterle, O. Painter, and D. E. Chang, New J. Phys. {\bf 21}, 025003 (2019).
\bibitem{Jen2020_subradiance} H. H. Jen, M.-S. Chang, G.-D. Lin, and Y.-C. Chen, Phys. Rev. A {\bf 101}, 023830 (2020).
\bibitem{Jen2020_PRR} H. H. Jen, Phys. Rev. Research {\bf 2}, 013097 (2020).
\bibitem{Jen2020_disorder} H. H. Jen, Phys. Rev. A {\bf 102}, 043525 (2020). 
\bibitem{Jen2021_bound} H. H. Jen, Phys. Rev. A {\bf 103}, 063711 (2021).
\bibitem{Zanner2022} M. Zanner, T. Orell, C. M. F. Schneider, R. Albert, S. Oleschko, M. L. Juan, M. Silveri, and G. Kirchmair, Nat. Phys. {\bf 18}, 538 (2022).
\bibitem{Pennetta2022} R. Pennetta, M. Blaha, A. Johnson, D. Lechner, P. Schneeweiss, J. Volz, and A. Rauschenbeutel, Phys. Rev. Lett. {\bf 128}, 073601 (2022).
\bibitem{Jen2022_correlation} H. H. Jen, Phys. Rev. A {\bf 105}, 023717 (2022). 
\bibitem{Pennetta2022_2} R. Pennetta, D. Lechner, M. Blaha , A. Rauschenbeutel, P. Schneeweiss, and J. Volz, Phys. Rev. Lett. {\bf 128}, 203601 (2022).
%%%%%%%%%%%%%%%%%%%%%%%%%%%%%%%%%% intro: topological waveguide-QED
\bibitem{Kim2021} E. Kim, X. Zhang, V. S. Ferreira, J. Banker, J. K. Iverson, A. Sipahigil, M. Bello, A. Gonz\'alez-Tudela, M. Mirhosseini, and O. Painter, Phys. Rev. X {\bf 11}, 011015 (2021).% topological waveguide-QED
%%%%%%%%%%%%%%%% free-space chiral QO
\bibitem{Grankin2018} A. Grankin, P. O. Guimond, D. V. Vasilyev, B. Vermersch, and P. Zoller, Phys. Rev. A {\bf 98}, 043825 (2018).
%%%%%%%%%%%%%%%%%%%%%%%%%%%%%%%%%%% intro: collective spin-phonon interactions
\bibitem{Shankar2019} A. Shankar, E. Jordan, K. A. Gilmore, A. Safavi-Naini, J. J. Bollinger, and M. J. Holland, Phys. Rev. A {\bf 99}, 023409 (2019).
%%%%%%%%%%%%%%%%%%%%%%%%%%%%%%%%%%% intro:ions QC
\bibitem{Pino2021} J. M. Pino, J. M. Dreiling, C. Figgatt, J. P. Gaebler, S. A. Moses, M. S. Allman, C. H. Baldwin, M. Foss-Feig, D. Hayes, K. Mayer, {\it et al.}, Nature {\bf 592}, 209 (2021).
\bibitem{Leibfried2003} D. Leibfried, R. Blatt, C. Monroe, and D. Wineland, Rev. Mod. Phys. {\bf 75}, 281 (2003).
\bibitem{Cirac1995} J. I. Cirac and P. Zoller, Phys. Rev. Lett. {\bf 74}, 4091 (1995).
\bibitem{Kielpinski2002} D. Kielpinski, C. Monroe, and D. J. Wineland, Nature {\bf 417}, 709 (2002).
\bibitem{Shen2020} Y.-C. Shen and G.-D. Lin, New J. Phys. {\bf 22}, 053032 (2020).
%%%%%%%%%%%%%%%%%%%%%%%%%%%%%%%%%%% intro:ions QS
\bibitem{Buluta2009} I. Buluta and F. Nori, Science {\bf 326}, 108 (2009).
\bibitem{Lanyon2011} B. P. Lanyon, C. Hempel, D. Nigg, M. Müller, R. Gerritsma, F. Zähringer,
P. Schindler, J. T. Barreiro, M. Rambach, G. Kirchmair, {\it et al.}, Science {\bf 334}, 57 (2011).
%%%%%%%%%%%%%%%%%%%%%%%%%%%%%%%%%%% ultracold atoms/molecules.
\bibitem{Hsiao2018} Y.-F. Hsiao, Y.-J. Lin, and Y.-C. Chen, Phys. Rev. A {\bf 98}, 033419 (2018).
\bibitem{Anderegg2018} L. Anderegg, B. L. Augenbraun, Y. Bao, S. Burchesky, L. W.
Cheuk, W. Ketterle, and J. M. Doyle, Nat. Phys. {\bf 14}, 890 (2018).
%%%%%%%%%%%%%%%%%%%%%%%%%%%%%%%%%%% EIT
\bibitem{Fleischhauer2005} M. Fleischhauer, A. Imamoglu, and J. P. Marangos, Rev. Mod. Phys. {\bf 77}, 633 (2005). 
%%%%%%%%%%%%%%%%%%%%%%%%%%%%%%%%%%% time dynamics
\bibitem{Roos2000_2} C. Roos, {\it Controlling the quantum state of trapped ions}, PhD thesis (University of Innsbruck, 2000).
%%%%%%%%%%%%%%%%%%%%%%%%%%%%%%%%%%% sideband
\bibitem{Chen2022} C.-C. Chen, Y.-C. Wang, C.-C. Wang, and H. H. Jen, arXiv:2203.00877.
\end{thebibliography}
\end{document}